# CLASSIFICATION OF SMART ENVIRONMENT SCENARIOS IN COMBINATION WITH A HUMAN-WEARABLE-ENVIRONMENT-COMMUNICATION USING WIRELESS CONNECTIVITY

Kristof Friess[1] and Prof. Dr. Dr. h.c. Volker Herwig[2]

[1]Department of Applied Computer Science, University of Applied Science Erfurt,
Erfurt, Germany
kristof.friess@fh-erfurt.de

[2]Department of Applied Computer Science, University of Applied Science Erfurt,
Erfurt, Germany
volker.herwig@fh-erfurt.de

## ABSTRACT

*The development of computer technology has been rapid. Not so long ago, the first computer was developed which was large and bulky. Now, the latest generation of smartphones has a calculation power, which would have been considered those of supercomputers in 1990. For a smart environment, the person recognition and re-recognition is an important topic. The distribution of new technologies like wearable computing is a new approach to the field of person recognition and re-recognition. This article lays out the idea of identifying and re-identifying wearable computing devices by listening to their wireless communication connectivity like Wi-Fi and Bluetooth and building a classification of interaction scenarios for the combination of human-wearable-environment.*

## 1. INTRODUCTION

With the growing market of worn computer systems like smartphones and smartwatches, in short wearables, the possible interaction between human and computer has changed. In a short time, also the interaction between human, computer and the environment will change. There are unlimited use cases where a human uses the computer as an assistant for filtering data, processing information or storing context information. Now and in the near future, there are a lot of use cases coming up, where not the device itself helps the human to become smarter, but the environment, based on the knowledge about the human, acts smartly. What those scenarios are and how they can be used is the core contribution of this paper. The base technology to implement those scenarios are technologies to recognize humans in the environments.

After an introduction to the state of human recognition, an approach using the wearable computing devices as a tag for smart environments is given. Then, Smart Environments are defined, meaning what are Smart Environments and what is done with a Smart Environment. On the basis of this definition, the interaction between the human and the wearable is enhanced by the interaction with the environment. This extension represents the flow of information between human-wearable and the environment and forms the basis for the classification of Smart Environments, which is developed afterwards.

## 2. HUMAN RECOGNITION

### 2.1 Established Recognition Techniques

The application field of human recognition and re-recognition is already there. There are many situations where video surveillance detects movements of humans for security reasons, for examples in factories, banks, airports, or country borders. Also at huge events, video surveillance is used for monitoring individuals and groups to detect panic situations, timely. But all those systems bring no additional benefit, apart from security, to the human, who is being recognized. The detection quality using video analysis and live images suffers, when people in the pictures are overlapping or are very small. When the overlap increases (more people), or people get smaller (because of the distance) in the image, then negative-positive results will be much higher [1] [2]. To re-recognize an object, in this case a human, features of the human (for example clothing) are needed. The required characteristics used for re-recognition can only be found at the human himself [2] [3, 4]. So, if the human changes his clothes, he can no longer be identified as the same human as before.

Another possible way of identifying humans is that every human inside the environment in focus is wearing a tag (e.g. infrared sender), so that each person can be detected. A problem is the sharing of the tag [4]. An alternative for this is the manipulation of the environment with sensors, like pressure sensors inside the ground panel of a floor (e.g. "active Floor" [6]). So, the environment is only knowing there is someone inside the floor, but not who, and also it does not know whether it is a returning human (re-recognition) [6].

The use of a tag, however, appears promising especially when considering the development of the spread of wearables [6] [7] [8]. Wearables, such as a smartphone, usually have wireless communication technology (Wi-Fi, Bluetooth). These technologies can be used to recognize people within an environment by different identification features (such as the network address) [4] [10]. Since most people carry a wearable, the equipment and a smart environment can be realized with simple means.

### 2.2. Recognition using Wearable Computing

The first wearables were developed more than 1000 years ago. In 1268, the philosopher Roger Bacon wrote about lenses for optical use – the glasses. This is the earliest mention of eyeglasses and the very first step towards wearables [11] [12]. Nearly 400 years later, an abacus ring appeared that was small and portable, and could be used as a calculator [12]. In the following years, more and more wearables were developed, often with mechanical technology such as small pocket watches. Then, in 1966 the first wearable computer was developed. Ed Thorp and Claude Shannon created a portable computer to predict the position of the ball in the casino game roulette [6] [7]. By now, the market for wearable computing has grown significantly and continues to grow [6] [7].

The question whether a computer is a wearable or not is answered by the pioneers in wearable computing. A definition of wearable computing comes from Thad Starner. He defines the wearable computer as an extension of the human body with passive use, similar to riding a bike. The idea is a very close connection of humans and computers, which allows a new way of perception that would not be possible without a computer. The computer is closely connected to the human and needs to make the same experience as the user. The computer learns from the user's environment and how he acts in certain situations. As the computer learns this, its ability in assisting the user increases [13].

Another definition of wearable computing is presented by Bradley J. Rhodes in 1997 in his article about wearables as remembrance agents [14]. He describes wearables based on the characteristics – portable while operational, hands-free use, sensors (GPS, camera) not only for input, proactive (getting in touch with the user), and many others. According to Bradley, a wearable computer should have many of those characteristics. He tries to show a clear distinction between a Palmtop PC and a wearable computing.

Next to Rhodes, Steve Mann published his definition of wearable computing at the International Conference on Wearable Computing ICWC-98 in 1998 [15]. Early on, he started to experiment with enhancements to his body. For example, he packed an 8-bit microprocessor 6502 in a backpack to control photo graphics equipment in 1981. He defined these devices as a new form of interaction between humans and computers. The computers are small, worn on the body, the system is always running, always ready, and always available. Those are the biggest differences to PDAs, handheld devices, and laptops. This new form of permanent availability and a given consistency of the user interface leads to new synergies between humans and computers [15] [16]. In detail, he sub-classifies wearables into three main characteristics (constancy, augmentation, mediation), which refer to the execution, and into six basic properties (un-monopolizing, unrestrictive, observable, controllable, attentive, communicative) in communication between humans and computers [15].

Looking at the different definitions of wearable computing, a clear concept reveals: Wearables are devices which should be integrated into our daily lives. They are always with us, and monitor us and the environment using sensors. The wearables are intended to assist people and support them secondarily (inform, analyze) while they are performing a primary activity. For that reason, wearables are a perfect basis for a new approach in personal recognition and re-recognition and in setting up some new concepts for smart environment.

The current methods of human recognition and re-recognition allow only limited automation. That can be explained by weaknesses in the algorithms [2]. People detection using video analysis and live images suffers from the challenge that people are overlapping and very small. As already stated before, the re-detection is challenged by the unique identification like clothes.

To recognize people across multiple cameras and over a longer period of time, an additional source of information is required, that is less likely to change, such as clothing. By considering modern technology like smartphones, which is already used by more than 49 million people [8] in Germany, there is a recognition feature that does not change every day.

Due to the high frequency of use, almost 75% of smartphone owners take their device with them [21] [22]. That technology is a good starting point to explore new approaches of recognition. In addition to the smartphone, the wearable computing market continues to grow as well [6] [7]. The devices range includes simple fitness trackers, smartwatches, as well as head-up displays like Google Glasses.

An extensive examination of the currently marketed wearables, done by the author, identified that most units have a wireless communication interface. In addition to GSM / EDGE / 3G and LTE, the most common technology is Wi-Fi, followed by Bluetooth and Bluetooth Low Energy. Other communication interfaces such as ANT + and NFC could also be identified, but currently the relevance and the distribution of ANT + is not as high as of Bluetooth and Wi-Fi. Furthermore, the NFC range is too short for widespread recognition.

Human recognition using Wi-Fi with a non-active connection to a hot spot and an unchanged hardware already has some experimental progress. This is made possible by periodically sent probes of the active Wi-Fi modules [4]. Therefore, wearables can be detected and optionally localized without being actively connected to a Wi-Fi [4].For example, a smartphone sends a Wi-Fi sample to detect access points close-by every 15 to 60 seconds, depending on the battery state [10]. This type of access point detection is used in multiple studies. In a high traffic street, 7,000 devices could be detected within a time frame of 12 hours [10]. In another study, around 160,000 devices could be detected during three hours [4, 19, 19]. A company in London was able to recognize people using this technology for several weeks. They could identify more than 530,000 smartphones within two weeks [23]. Further experiments have shown that the recognition using Wi-Fi probes is promising as well [24] [25] [23].

In addition to the already existing tests with Wi-Fi, there are other approaches using Bluetooth. Bluetooth was created in 1998 with the merge of several companies under the name Bluetooth Special Interest Group (SIG) [27] [28]. In the following year, version 1.0 was released. Bluetooth is now installed in most phones and is therefore another possibility for mobile positioning and identification of mobile devices. With the release of Bluetooth 4.0 standards in 2010, the new low energy technology was created, which makes it possible to maintain a radio link with low power consumption for a longer time [27] [29] [27]. Using this specification, a company enabled the opportunity for recognizing objects as well as people using small beacon modules. Using the beacons, it is also possible, according to a previously performed calibration, to locate people with a smartphone in premises [31] [32].

It has been shown that several approaches for human recognition and re-recognition using radio wave technologies are available. Almost all of the previously revealed methods use unmodified hardware carried by the people. For this concept, we focus on Wi-Fi and Bluetooth connectivity. Because with the current state of the art knowledge, we know that these two wireless technologies have the possibility to re-identify a user using the MAC or IDs of devices, so this are the best enabling technology for a smart environment. With the scenarios described above, the technology used has to be evaluated in regards to sources of error and the ability for application.

## 3. SMART ENVIRONMENT BASED ON WEARABLE COMPUTING

As early as the early 1990s, Mark Weiser and his colleagues came to the conclusion that the PCs of the world and the change to the laptop is only an intermediate step in the transformation to the smart environment [33]. The idea of Weiser is that the computer disappears in the background of the environment and is thus not the main focus of the utilizing human being. On the contrary, man should concentrate on the fellow human beings and his activities [33] [19]. The merging of the environment with computers, the concept that computers can appear in various forms at all times, is summarized under the term ubiquitous computing. Ubiquitous computing (or brief ubicomp) is the idea of a "physical world richly and invisibly interwoven with sensors, actuators, displays, and computational elements, embedded seamlessly in the everyday objects of our lives and connected through a continuous network" [19], from this idea the Smart Environment developed. smart environment, or even Smart Space, is a "region of the real world that is extensively equipped with sensors, actuators and computing components" [31] [32]. So an environment with many sensors, motors and computer components as it has already described in [33]. The described environment opens up new possibilities for interaction and usability. It is a combination of several heterogeneous systems "systems and networks, between systems and systems, and between systems and people" [32].

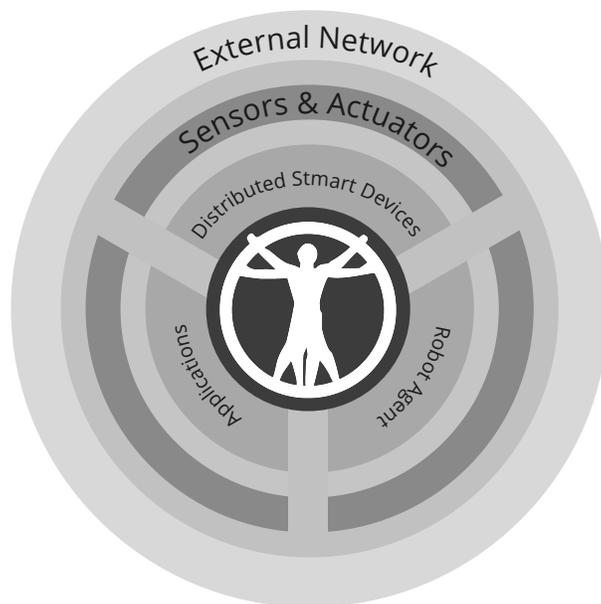

Figure 1. Schematic representation of smart environments [37]

Cook provides a similar description: Smart Environments are a "…small world where all kinds of smart devices are continuously working to make inhabitants' lives more comfortable…" [33]. In this smart or intelligent environment, it is possible to autonomously assess the nature and the individuals and to make decisions from this knowledge. The goal is to improve comfort for the individual and the experience [33]. A schematic view on smart environment is given in Figure 1.

Based on these smart environment descriptions, the following definition is used for this article:

Smart Environments are environments with a significant amount of computer components, motors, and sensors. These components, however, unlike the personal computer, are fused with the environment and thus invisible to the individuals who are in it. The environment is connected via an internal and external communication network and knows how to deal with the data it raises. On the basis of the information gathered, the environment can perform tasks that help individuals and increase comfort.

Examples of Smart Environments are ubiquitous. Smart Smarties, SmartHouses, SmartHealths, SmartLaboratories, SmartLaboratories and SmartRooms are a way of realizing smart environments [33] [32] [34].

Already in 1989, there were first experiments to realize a smart environment. At the Olivetti Research Labs/ University of Cambridge, people were able to be located in rooms and also in the building using an "Active Badge". This was possible because the active mark periodically transmitted an infrared signal to the sensors positioned in the room [39]. However, it could only be ascertained whether a person is in a room, but not where. In order to improve localization, a new active marker was used in 1995. With the help of transmitted ultrasonic waves, it was now possible to determine the position of a person with an accuracy of up to 3cm in the room. The name for the new marker was "Active Bat". This is due to the fact that a bat uses a similar system for position detection [39]. In addition to the identification of persons in rooms using markers, there are also systems that function without a marker. A good example is the Smart Floor. By installing pressure sensors in the floor slabs of a hallway, persons walking across could be recognized [36].

It is clear that a smart environment can only be created, if the environment knows that there is a person, and in the best case, which person is in it. The listed examples are only a small set of actual implementations, but they illustrate the need for manipulation. Manipulation means that the environment must be equipped with sensors in such a way that they can recognize people and / or that the persons must be "manipulated" by carrying a marker. The latter can be supported by the wearing of a wearable.

Because by carrying smartphones, smartwatches and other portable computers with radio technologies, there is an "active tag", a marker for the environment. The fact that most people already carry an active marker every day promotes the development of a smart environment with significantly fewer sensors. This is because the environment can use the wearables' functional technologies within a certain range for the communication or identification of the person and recognize, locate and identify the person. With this information and the information that the environment receives from the other sensors and components, the support and thus the extension of the experiences of the persons within the environment can begin.

When considering the flow of information between humans and computers as defined by Steve Mann, the wearable encompasses the human being and a permanent interaction between man and wearable is possible [15]. This flow of information is now extended and allows the wearable, which is to be understood as a protection for the privacy as well as the extension of the senses [15], a communication with the environment. For this information flow to work, it is necessary that the wearables are equipped with an active wireless technology such as Bluetooth or Wi-Fi, for data exchange.

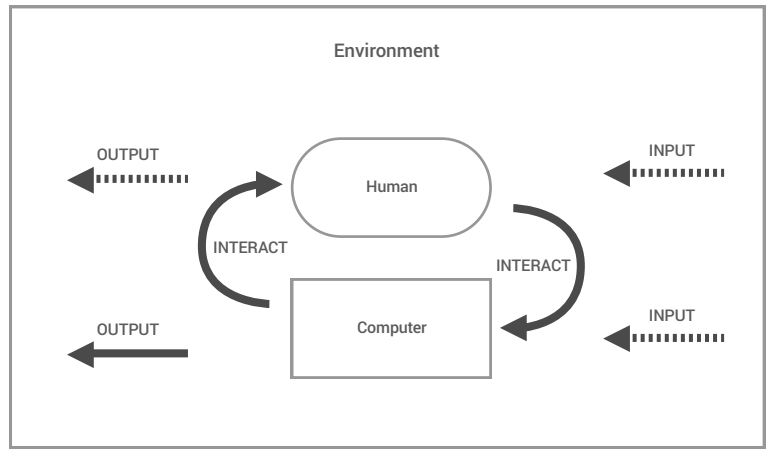

Figure 2. Human-Wearable-Environment information flow

The interaction between human-computer defined by Steve Mann has been expanded in the Figure 2 by the information flow (in- / output) between the human computer environment. This shows how the communication between the environment and the human being and the environment and the wearable functions. The wearable always has an active functionality and always sends an output that identifies the person. Since this identification must be activated continuously, a connection with the environment in Figure 2 is indicated by a continuous arrow. The environment itself can provide information to people or the computer / wearable, either via radio technologies, sensors, screens or audio. However, this is not always necessary and is therefore shown in the figure with a dashed arrow. The same is true for the human. He can be a supplier of information, for example, by talking, pressing a switch or being captured by a camera.

## 4. CLASSIFICATION OF SMART ENVIRONMENT SCENARIOS

A second factor plays an important role in considering the flow of information between human-wearable and the environment. This is the way communication / interaction with the environment takes place and how data is exchanged. In order to describe this communication and interaction in a more general way, a classification of scenarios takes place. When considering different scenarios for Smart Environments, it could be revealed that these differ in the provision of the information of human-wearable, in the nature of the interaction which originates from the environment and in the kind of the feedback.

It was previously defined that the wearables have a passive communication based on their radio technologies with the environment. The wearable and also the human being get nothing in return from this kind of data transmission, therefore this is called a *Passive Tag*. It is different when the wearable uses the radio connection to transmit information to the environment – *Active Tag*.

**Passive Tag** – The human being and the wearable, too, are not aware that they are perceived by the environment. The wearable is detected by the active radio technology (e.g. Wi-Fi, Bluetooth). Based on this, the human-wearable can be identified. Depending on how the environment processes the information about the human, it can learn the behavior of human beings and support them in the future through interactions.

**Active Tag** – The human being wants the environment to perceive him and provides additional information about the wearable. This information will personalize the people towards the environment. For example, people could share the shopping list with the environment and this responds with an optimized shopping route. Further information can be people's interests, such as politics, music taste and sports. All this information is used to optimize the interaction and personalization of the feedback to the people, for example in the form of tailored advertising on screens, adapted room temperature or music.

With the information that the environment receives from the active or passive tags, it can make decisions for the interaction, how the environment reacts to the human-wearable. In considering how a smart environment interacts with the human-wearable combination, three different interaction possibilities can be classified.

**Direct Interaction** – There will be a directly returned feedback to the human-wearable. This may be accomplished by performing various tasks, such as turning lights on / off, adjusting room temperature or tracking the person with the favorite music when walking through the premises of a building. Ultimately, the environment directly affects and supports the people.

**Indirect Interaction** – The human-wearable does not get any direct feedback from the environment. The environment uses the information for analysis. The information obtained (e.g. the number of persons, movement currents, actions of the persons) are used to learn future actions / decisions (e.g. create a new direct interaction automatic open and close windows or recognizing dangerous situations).

**No Interaction** - The human-wearable gets nothing from the environment, all the information that the environment captures are used for analysis. This is the case, for example, when analyzing a trade fair. It is determined how many people have attended a stand. This information serves purely the statistics and is not returned to the human-wearable in any form.

The smart environment is designed to improve the user experience, to expand the comfort of the people and to support them in the everyday life from the background. Not always, as defined above, direct communication with the human-wearable occurs. However, the surrounding environment collects data. A distinction in which form the environment processes the data and deduces actions from it is decisive. There are four different types of feedback.

**Navigation** – All scenarios supporting the control and navigation of human-wearable within the environment. An example is a person who has just entered a train. This person does not know where there are still free seats, but the train does. Therefore, direct control of this person is done using the wearables or monitors to find a free seat. A similar scenario is a parking garage. The surrounding area would directly navigate to the next free space.

**Content** – All scenarios in which the environment uses the collected information about human-wearable to provide optimized content for it. This is the case, for example, with tailored advertising and optimized messages. The human being gets the information he is interested in.

**Observation** – Identifies all scenarios in which the environment is used as a pure observer. This is the case, for example, when analyzing the number of visitors in buildings or at trade fairs.

**Trigger** – Identifies all scenarios in which the environment triggers an action for human-wearable, for example open a door or turning on/off the light.

In order to understand why this classification has been decided, some scenarios will follow, which illustrate the respective information flows and interactions. The scenarios constructed in this way are fictive and their realization and meaningfulness should not be evaluated. It is important, however, that there are people in it who carry a wearable with active radio technology (such as Bluetooth or Wi-Fi).

**Smart Train: My Seat**

At the main station of Erfurt is Sebastian S., he has an electronic ticket from Erfurt to Berlin on his smartphone and is waiting for his ICE. While waiting for his train, his smartphone informs him that he is currently in the wrong area. His reserved seat is in the area E. After the train has entered, Sebastian S. enters the train, on a monitor directly in front of him is indicated that his seat 32F is on the left side. Sebastian S. follows this instruction and quickly finds his seat by a flashing light.

This scenario is an active tag. Because the environment got the information from the electronic ticket and could use this information to carry out a supporting navigation. Sebastian S. was localized and navigated on the basis of his smartphone. The interaction was direct with the human-wearable.

*Classification:* active tag, direct interaction, navigation

**Smart Train: Detect a free Seat**

At the main station of Dresden is Lucas L., he has an electronic ticket from Dresden to Munich on his smartphone and is waiting for his ICE. While waiting for his train, a monitor informs him that there are still lots of free seats in the Wagon 21. The wagon 21 stops in area B. After the train has arrived, Lucas enters. The monitor in the entrance area informs him that there are still a lot of free seats on the right and recommends to go to the right.

This scenario is a passive tag. Because the environment only gets the information that there is a person with a wearable in it. On the basis of this information and further information about the current characteristics in the incoming train, navigation could be carried out by means of environmental elements. Lucas L. was localized and navigated on the basis of his smartphone. The interaction was direct with the human-wearable.

*Classification:* passive tag, direct interaction, navigation

**Optimized Advertisement**

Maria M. is at the fair CEBIT. She visits different stands longer, others shorter. Certain booths she does not visit at all. As she moves around the trade fair, the monitors display ads that relate to topics based on the booths Maria M. has already visited.

In this example, in addition to indirect and direct interaction, two different types of feedback are also shown. First, it is assumed that Maria M. does not provide any information about her wearable. That is, it has a passive tag. In order to personalize the advertising for Maria M., the surroundings must learn about the interests of Maria M. by means of the up-and-coming days at the respective stands. This learning is positioned in the indirect interaction with human wearable. If there is sufficient knowledge, the environment can derive action decisions from this – direct interaction. Maria M. also has the possibility to wear an active tag. In doing so, they pass on information about their interests to the environment and the environment can start directly with the interaction. In the nature of the feedback, there is initially the presentation of content and with the recommendation of further fair stands a navigation takes place.

*Classification:* passive / active tag, indirect / direct interaction, content / navigation

**People-Flow**

The passenger flows are measured in a shopping center. The measurements serve to analyze the area in which advertising is best placed: where visitors are most likely to stay and which areas are not visited at all.

This is a scenario where visitors are tracked and analyzed with their passive tag. This is done by observation only. The environment does not derive any action decisions from this information.

*Classification:* passive day, no interaction, observation

**Smart Buildings**
Every morning at around 8 o'clock, Katrin K. arrives at the office. At first, she switches on the light in the entrance area, then in the corridor and in her room.

One day, when Katrin K. arrives, the light turns on automatically in the entrance area, then in the corridor and in her room.

In this scenario, it is obvious that the environment can derive its own action by means of the persons and the actions. The environment is first to be classified as an indirect interaction. After the environment has learned how to support the human-wearable, it is a direct interactor.

*Classification:* passive day, indirect / direct interaction, observation / trigger

These are only a small part of scenarios that can be described with this classification. During the review, further scenarios were concieved and found that this classification is useful for building further possible scenarios and ideas for Smart Environments.

## 5. CONCLUSIONS

Wearable computing is a good system for supporting Smart Environments. Through the focus of wearable computing which learns from humans and supports them in everyday life, the wearable (for example, a smartphone) becomes an assistant. The integrated wireless technologies provide a data exchange interface, which can be used by other wearables or environments for the information exchange.

In the case of a human wearable environment, it is assumed that the wearable is always worn by humans. The interaction is, as Steve Mann defined, encapsulated and continuous. That is, the wearable has the possibility to demand the attention of humans and to interact with them. The human being always has the possibility to interact with his wearable. The environment surrounds people with their wearable. A continuous flow of information from the wearable to the environment takes place via the radio technology. The environment can decide how and whether it interacts with the human-wearable and / or exchanges information.

The possibility that the environment does not necessarily have an interaction / information flow with the human wearable is the basis for the creation of a classification of interaction scenarios between humans and the environment. It was possible to identify a wearable as an active or passive information provider in the environment. A wearable with active radio connection is a passive tag, the radio connection and an identifier of the device can idenitfy the person within the environment. If the wearable additionally provides information about the person, this is referred to as an active tag. Whether active or passive, the environment can learn from the actions of the person and derive their own actions from it. This learning is referred to as indirect interaction because the environment first processes the data, and it is only when the environment develops actions (switching on light) that direct interaction occurs. However, it may also be that the environment only functions as an information collector, that is, there is no interaction. Depending on which actions the environment chooses to interact with the human-wearable, the feedback differs. If the environment switches on the light when entering the room, then this is a trigger. If the environment responds with, for example, advertising on monitors, this is a substantive feedback. If the environment tries to control the people in it by means of directional indications on the monitors, this is a navigation feedback. Or as briefly described above, the environment analyzes the people within without feedback and only observes them. Thus, different scenarios can be classified with the distinction into active or passive tags, the type of interaction (direct, indirect or none), and the type of feedback (content, navigation, observation, trigger).

## REFERENCES

[1] E. Corvee, S. Bak and F. Bremond, "People detection and re-identification for multi surveillance cameras," VISAPP - International Conference on Computer Vision Theory and Applications - 2012, 2012.

[2] P. Dollar, C. Wojek, B. Schiele and P. Perona, "Pedestrian Detection: An Evaluation of the State of the Art," IEEE Transactions on Pattern Analysis and Machine Intelligence, 2012.

[3] F. J. Baek, M. K. Islam and Joong-Hwan, "Person Detection, Re-identification and Tracking Using Spatio-Color-based Model for Non-Overlapping Multi-Camera Surveillance Systems," Smart Computing Review, vol. 2, no. 1, p. 42, 2012.

[4] M. V. Barbera, A. Epasto, A. Mei, V. C. Perta and J. Stefa, "Signals from the Crowd: Uncovering Social Relationships Through Smartphone Probes," in Proceedings of the 2013 Conference on Internet Measurement Conference, New York, NY, USA, 2013.

[5] S. Poslad, UbiquitousComputing: Smart Devices, Environments and Interactions, London: John Wiley & Sons, Ltd., 2009, p. 42.

[6]  M. Addlesee, A. Jones, F. Livesey and F. Samaria, "The ORL Active Floor," IEEE Personal Communications, 1997.

[7]  GfK, "Absatz von Smartwatches in Deutschland in den Jahren 2013 bis 2015," 2015. [Online]. Available: http://de.statista.com/statistik/daten/studie/459093/umfrage/absatz-von-smartwatches-in-deutschland/. [Accessed 10 03 2016].

[8]  IDC, "Prognose zum Absatz von Wearables weltweit von 2014 bis 2020 (in Millionen Stück)," 01 01 2016. [Online]. Available: http://de.statista.com/statistik/daten/studie/417580/umfrage/prognose-zum-absatz-von-wearables/. [Accessed 29 03 2016].

[9]  Statista, "Anzahl der Smartphone-Nutzer in Deutschland in den Jahren 2009 bis 2016 (in Millionen)," [Online]. Available: http://de.statista.com/statistik/daten/studie/198959/umfrage/anzahl-der-smartphonenutzer-in-deutschland-seit-2010/. [Accessed 26 07 2016].

[10] A. B. M. Musa and J. Eriksson, "Tracking Unmodified Smartphones Using Wi-fi Monitors," in Proceedings of the 10th ACM Conference on Embedded Network Sensor Systems, New York, NY, USA, 2012.

[11] B. Rhodes, "A brief history of wearable computing," 2001. [Online]. Available: https://www.media.mit.edu/wearables/lizzy/timeline.html. [Accessed 29 03 2016].

[12] L. d. Medici, "The History Of Wearable Technology - Past, Present And Future," 15 11 2015. [Online]. Available: https://wtvox.com/featured-news/history-of-wearable-technology-2/. [Accessed 29 03 2016].

[13] T. Starner, "The challenges of wearable computing: Part 1," IEEE Micro, vol. 21, no. 4, pp. 44 - 52, Juli-August 2001.

[14] B. J. Rhodes, "The wearable remembrance agent: a system for augmented memory," Wearable Computers, 1997. Digest of Papers., First International Symposium on, pp. 123-128, 1997.

[15] S. Mann, "Definition of "Wearable Computer"," 1999. [Online]. Available: http://wearcam.org/wearcompdef.html. [Accessed 10 03 2016].

[16] S. Mann, "WEARABLE COMPUTING as means for PERSONAL EMPOWERMENT," 25 08 1998. [Online]. Available: http://wearcam.org/icwckeynote.html. [Accessed 10 03 2016].

[17] BITKOM, "Das Handy als ständiger Begleiter," 2015. [Online]. Available: http://www.bitkom.org/de/presse/64046_77337.aspx. [Accessed 01 03 2015].

[18] TNS Infratest Mobile Club 2013, "Smartphone-Nutzung und ihre Einsatzorte," 2015. [Online]. Available: http://www.tns-infratest.com/Presse/pdf/Presse/2013_05_06_TNS_Infrastest_Mobile_Club_Mobile-Landscape_Charts.pdf. [Accessed 02 03 2015].

[19] M. Weiser, R. Gold and J. S. Brown, "The origins of ubiquitous computing research at PARC in the late1980s," IBM SYSTEMS JOURNA, vol. 38, no. 4, 1999.

[20] K. a. Shubber, "Presence Orb uses Wi-Fi to detect if buses and bars are full," 01 04 2015. [Online]. Available: http://www.wired.co.uk/news/archive/2014-04/22/presence-orb. [Accessed 20 09 2015].

[21] R. Lim, "Tracking Smartphones Using Low-power Sensor Nodes," in Proceedings of the 11th ACM Conference on Embedded Networked Sensor Systems, New York, NY, USA, 2013.

[22] J. Freudiger, "How Talkative is Your Mobile Device?: An Experimental Study of Wi-Fi Probe Requests," in Proceedings of the 8th ACM Conference on Security & Privacy in Wireless and Mobile Networks, New York, NY, USA, 2015.

[23] K. Li, C. Yuen and S. Kanhere, "SenseFlow: An Experimental Study of People Tracking," in Proceedings of the 6th ACM Workshop on Real World Wireless Sensor Networks, New York, NY, USA, 2015.

[24] Bluetooth, "History of the Bluetooth Special Interest Group," 04 01 2015. [Online]. Available: http://www.bluetooth.com/Pages/History-of-Bluetooth.aspx. [Accessed 01 03 2015].

[25] M. Sauter, Communication Systems for the Mobile Information Society, John Wiley & Sons Ltd.

[26] Bluetooth, "Bluetooth Smart Technology: Powering the Internet of Things," 2015. [Online]. Available: http://www.bluetooth.com/Pages/Bluetooth-Smart.aspx. [Accessed 04 03 2015].

[27] K. T. Davidson, C. Cuffi, Akiba and Robert, Getting Started with Bluetooth Low Energy, Media, O'Reilly, 2014.

[28] Estimote, "Indoor Location," 2015. [Online]. Available: http://estimote.com/indoor/. [Accessed 06 03 2015].

[29] Apple, "Understanding iBeacon," 2015. [Online]. Available: https://support.apple.com/en-us/HT202880. [Accessed 06 03 2015].

[30] M. Weiser, "The Computer for the 21st Century," SIGMOBILE Mob. Comput. Commun. Rev., vol. 3, no. 3, pp. 3-11, 1999.

[31] P. A. Nixon, W. Wagealla, C. English and S. Terzis, Security, Privacy and Trust Issues in Smart Environment, 2004.

[32] P. N. Dobson, G. Lacey and Simon, Managing Interactions in Smart Environments: 1st International Workshop on Managing Interactions in Smart Environments (MANSE'99), Dublin, December 1999, London: Springer-Verlag, 1999, pp. 1,3.

[33] D. J. C. Das and K. Sajal, Smart Environments: Technologies, Protocols, and Applications, John Wiley & Sons, Inc., 2005, pp. 3,4,9.

[34] F. Marquardt and A. M. Uhrmacher, "Evaluating AI Planning for Service Composition in Smart Environments," Proceedings of the 7th International Conference on Mobile and Ubiquitous Multimedia, vol. MUM '08, pp. 48-55, 2008.

[35] S. Poslad, UbiquitousComputing: Smart Devices, Environments and Interactions, John Wiley & Sons, Ltd., 2009, p. 42.

[36] M. Addlesee, A. Jones, F. Livesey and F. Samaria, "The ORL Active Floor," IEEE Personal Communications, 1997.

[37] FairControl, "Controlling von Live-Kommunikation," 2013. [Online]. Available: www.faircontrol.de. [Accessed 06 08 2013].

[38] GMM Gelszus Messe-Marktforschung, "GMM - Experten für mobile Marktforschung vor Ort," 01 01 2013. [Online]. Available: http://www.gmm-marktforschung.de/. [Accessed 26 08 2013].

[39] drinktec, "Lead Tracking," 01 01 2013. [Online]. Available: http://www.drinktec.com/en/Home/ForExhibitors/exhibitor-services/leadtracking. [Accessed 10 08 2013].

[40] Statista, "Anzahl der Smartphone-Nutzer in Deutschland in den Jahren 2009 bis 2015 (in Millionen)," ComSource, 24 09 2015. [Online]. Available: http://de.statista.com/statistik/daten/studie/198959/umfrage/anzahl-der-smartphonenutzer-in-deutschland-seit-2010/. [Accessed 24 09 2015].

[41] E. Corvee, S. Bak and F. Bremond, "People detection and re-identification for multi surveillance cameras," VISAPP - International Conference on Computer Vision Theory and Applications -2012, Februar 2012.

**Author**

Kristof Friess

Since 2016 – Doctoral Student, University of Applied Science Erfurt

2014 – Master of Science, University of Applied Science Erfurt

2011 – Bachelor of Science, University of Applied Science Erfurt

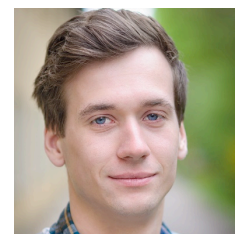

Volker Herwig

He is professor for business systems at the University for Applied Sciences in Erfurt, Germany. His focus is on mobil business systems and IT strategy. He worked 15 years in the industry in the USA and Germany.

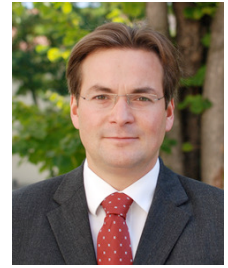